\documentclass[12pt,letterpaper]{article}   
\usepackage{osajnl} 
\begin{document}

\title{The optical performance of frequency selective bolometers}

\author{T.A.~Perera, T.P.~Downes, S.S.~Meyer, and T.M.~Crawford } \address{University
of Chicago, KICP, Chicago, IL 60637, U.S.A}

\author{E.S.~Cheng}
\address{Conceptual Analytics, Glenn Dale, MD 20769, U.S.A.}

\author{T.C.~Chen, D.A.~Cottingham, and E.H.~Sharp} \address{Global
Science and Technology, Greenbelt, MD 20771, U.S.A.}

\author{R.F.~Silverberg}
\address{NASA/GSFC, LASP, Greenbelt, MD 20771}

\author{F.M.~Finkbeiner and D.J.~Fixsen}
\address{SSAI, Greenbelt, MD 20706}

\author{D.W.~Logan and G.W.~Wilson} \address{University of
Massachusetts, Department of Astronomy, Amherst, MA 01003, U.S.A.}

\begin{abstract*}Frequency Selective Bolometers (FSBs) are a new type of detector for millimeter and sub-millimeter wavelengths that are transparent to all but a narrow range of frequencies as set by characteristics of the absorber itself.  Therefore, stacks of FSBs tuned to different frequencies provide a low-loss compact method for utilizing a large fraction of the light collected by a telescope.  Tests of prototype FSBs, described here, indicate that the absorption spectra are well predicted by models, that peak absolute absorption efficiencies of order 50\% are attainable, and that their out-of-band transmission is high. \\
\end{abstract*}


\maketitle


\section{Introduction}
\label{intro}

The millimeter and sub-millimeter wavelengths are a proven spectral
range for tracing the evolutionary history of our universe.  With
sources ranging from the cosmic microwave background, to high-redshift
starburst galaxies, to galaxy clusters, to cold dusty objects in our
own galaxy, this portion of the spectrum is uniquely fertile with
information on cosmological parameters, growth of structure from the
linear through the non-linear regime, star formation history and
galactic dynamics.  Advances in detector technology at millimeter and
sub-millimeter wavelengths have resulted in receiver sensitivities
approaching the background limit.  More recently, large format arrays
have been developed at these wavelengths for imaging from single-dish
telescopes (e. g. BOLOCAM, AzTEC, SHARC-2, SCUBA-2) and a new era of
high angular resolution ground-based observing will soon begin with
the first-light of the 50~m Large Millimeter Telescope (LMT) and the
Atacama Large Millimeter Array (ALMA) array.

Increasing detector count and field of view is not the only avenue for
improvement of today's photometers.  Typically there is a large
mismatch between the working bandwidth of telescopes and the bandwidth
of focal plane arrays.  We describe the concept and first
characterization of a new form of bolometric detector designed to
alleviate this inefficiency at millimeter and sub-millimeter
wavelengths.  Frequency Selective Bolometers (FSBs) facilitate a
dramatic improvement in the fraction of radiation utilized by a compact
and convenient focal plane array.

Multi-frequency photometric capabilities are currently achieved with
identical detectors behind dichroic beam-splitter setups or behind
different filters on an allocated focal plane.  These complex
arrangements make inefficient use of valuable focal plane area and/or
reduce optical efficiency with multiple chromatic splitters and
numerous reflections.  In addition, the focal plane optics can be
physically large, massive, and difficult to cool down to sub-Kelvin
temperatures and integrate into the tight constraints imposed on
orbital and sub-orbital missions.  FSB technology removes many of
these problems by incorporating a bolometer element into a resonant
structure to control the range of absorbing frequencies.  The key
advantage of the FSB is that radiation not absorbed by the bolometer
is transmitted with low loss through the device and passed on to
subsequent FSB elements tuned to other frequencies.\cite{kowitt96fsb}
A stack of FSBs, is therefore, a multi-spectral single pixel with an
angular resolution set by the lowest frequency detector.  Adjacent
stacks of FSBs can be efficiently close-packed into an array of
multi-color pixels populating the focal plane.

The SPEctral Energy Distribution (SPEED)
camera,\cite{WilsonSpeed2003,SilverbergSpeed04} scheduled for
commissioning in spring 2007, will be the first instrument to make use
of FSB technology.  SPEED will have four focal plane pixels, each
sensitive to four separate frequency bands centered around 148, 219,
271, and 361~GHz.  It will be used for photometry of distant dusty
galaxies previously found by surveys like Spitzer and SCUBA, and for
studying other millimeter-wave sources such as ultra luminous infrared
galaxies, the Sunyaev-Zeldovich effect in clusters, and galactic dust.
We present here the design and characterization of SPEED prototype
detectors.


\section{Optical design}
\label{design}

\begin{figure}[htbp]
\centerline{\includegraphics{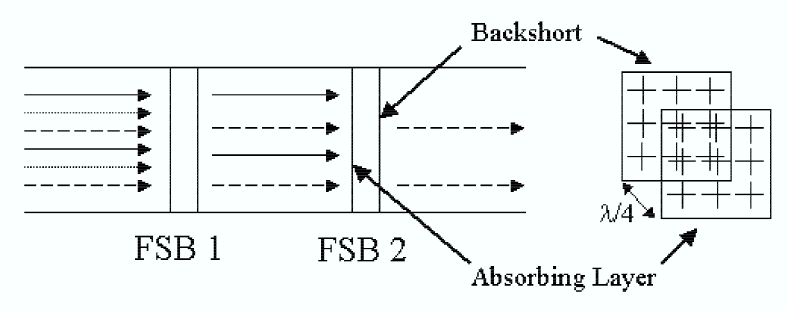}}
\caption{Cartoon of a stack of two FSBs.  Of the radiation entering
from the left, the component in the band of FSB~1 (dotted arrows) is
absorbed and radiation outside that band passes through the
detector. Of the remaining light, radiation that is within the band of
FSB~2 (solid arrows) is absorbed and all other radiation (dashed
arrows) continues down the stack until absorbed by an additional FSB
or by a terminating absorber. On the right side is a cartoon
illustrating the combination of absorbing layer and reflecting
backshort.}
\label{FSBStack}
\end{figure}

FSBs incorporate frequency selective surfaces as part of the detector
to define the absorption band. High out-of-band transmission allows
multiple, differently tuned FSBs to be stacked in series to create a
compact multi-color photometer.  Fig.~\ref{FSBStack} outlines the
concept of FSBs in a two color photometer.

\begin{figure}[htbp]
\centerline{\includegraphics{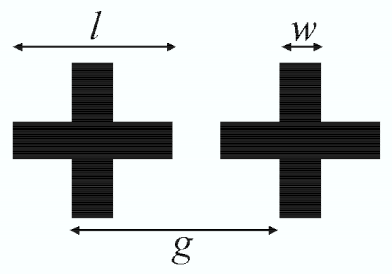}}
\caption{FSB surface geometry.} 
\label{FSBDiagram}
\end{figure}

The optical performance of an individual FSB is characterized by the
spectral profile of its absorption, the absolute efficiency of the
absorption, and its transparency to radiation outside the absorption
band.  The crossed-dipole geometry (see Fig.~\ref{FSBDiagram}) of the
FSBs described in this paper has been shown to produce a narrow
resonance with out-of-band transmission near unity.\cite{Munk00}  The
absorber and backshort geometries are described by three
parameters: the bar length, $l$, the bar width, $w$, and the
periodicity of the grid, $g$.  The resonant wavelength is near $2l$
and has a weak dependence on $w$.\cite{chase83}

Unlike frequency selective surfaces used strictly as reflective
filters, FSBs require efficient absorption of incident radiation.  The
frequency selective absorbers are made lossy to match the impedance of
free space and maximize coupling to plane wave radiation. Previous
studies of lossy frequency selective elements have demonstrated how
resistive periodic elements can efficiently absorb radiation within
narrow bands~\cite{Severin1956}.  The absorption of an FSB is
enhanced when paired with a similarly resonant but purely reflective
(lossless) backshort located one quarter wavelength behind the
absorber.  The backshort separation and other parameters that maximize
absorption at the chosen frequency fully constrain the spectral
profile of the absorption.  While both components of polarization are
absorbed by the FSBs discussed here, polarization sensitivity may be
achieved by replacing crosses in the metallic patterns with bars.

\subsection{HFSS modeling}
\label{hfss_modeling}

We numerically calculate the spectral response of potential FSB
designs to optimize the spectral shape and absorption efficiency.  We
use Ansoft's High Frequency Structure Simulator (HFSS)\cite{hfss} to
predict the transmission, reflection, and absorption of each detector.
While isolated crossed dipole frequency selective surfaces have been
described analytically with some success,\cite{pelton79} HFSS allows
modeling of realistic devices with substrate materials and
interactions between adjacent resonant meshes.

Each FSB is modeled as a unit cell with side lengths equal to the
periodicity of the grid and periodic boundary conditions on four sides
to simulate an infinite pattern of crosses.  Two perfectly matched
layers (PMLs)\cite{wu95,berenger96} on the top and bottom of the unit
cell terminate the incident wave and minimize edge effects. Inside the
unit cell are two aligned 2-D crosses representing an absorbing and a
backshort layer on substrates separated by a distance of
$\lambda_\mathrm{res}/4$ and normal to the plane wave excitation. The
spectral performance of the FSB is calculated by solving for the
transmission ($T$) and reflection ($R$) of incident plane waves at
frequencies where the unit cell model is valid ($c/2g < \nu <
c/g$).  Absorption is estimated as $1-R-T$.  These
models are used to optimize the spectral profile of all FSB designs
before fabrication.


\subsection{Physical description}

We have fabricated a set of FSBs with target resonant frequencies of
219 and 271~GHz. The devices were fabricated at the NASA Goddard Space
Flight Center (GSFC) using photolithographic techniques.

The absorber and backshort layers are built on a 0.5~$\mu$m thick film
of Si$_3$N$_4$ which is suspended from a Si frame. The size of the
suspended windows is approximately $11\times11$~mm. The backshort
cross pattern is formed of Au 200~nm thick to provide high
conductivity. The absorbing layer has Au 28~nm thick corresponding to
a sheet resistance of 1.6~$\Omega$/square, the value determined by
simulations to optimize absorption. Because currents are highest near
the edges of patterned features, a lift-off process with an overhang
on the resist is used to obtain sharp edges.  The crosses are arranged
to fill a 10~mm diameter light pipe centered on the element.
Table~\ref{FSBTable} gives the geometric properties of the crossed
dipole surfaces of each FSB.

To form an FSB, the bolometer and backshort are glued together
with a Si shim of thickness $\lambda_\mathrm{res}/4$.  The crosses on the two
layers must be registered to better than $\lambda/8$ for performance
that agrees with models. The aligning, therefore, is performed under a
microscope to adjust the lateral translation and rotation before the
glue sets.

The Si$_3$N$_4$ film is perforated so the absorbing layer is attached
to the frame by only four narrow legs.  A transition edge sensor (TES)
made from a Mo-Au bilayer is placed on the film outside the central
10~mm circle.\cite{FMFinkbeinerBilayer99,TCChenBilayer04}  The
electrical connections for the TES are provided by two leads that run
down the adjacent leg to wire-bond pads on the Si frame.  Two sets of
detectors were fabricated for the tests described here, one set with
Ti-Al-Ti leads, the other with Mo leads.  For redundancy, the
absorbing element has two TESs at opposite corners, but only one
operates during testing.  The absorbing layer of an FSB is shown in
Fig.~\ref{fsb_pics}.


\section{Thermal picture and readout scheme}
\label{g_model}

The TES is wired in parallel with a 6~m$\Omega$ ``shunt'' resistor
($R_b$) and biased into its transition with a known current through
the shunt-TES combination (see Fig.~\ref{TES_circuit}).  With normal
resistance between 100-200 m$\Omega$, TESs are essentially voltage
biased when not at the lower end of the transition.  Changes in
current through the TES ($\Delta I_\mathrm{TES}$) are measured with an
8$\times$1 time-division SQUID multiplexer.\cite{chervenak99}  A
first-stage SQUID and its pickup coil are also indicated in
Fig.~\ref{TES_circuit}.  The board carrying the SQUID multiplexer chip
and shunt resistors is attached to the $\sim$270~mK cold stage of the
cryostat.

The TES transition temperature (460~mK) and the absorption-band
dependent bolometer-to-bath thermal impedance were chosen to minimize
detector noise given typical temperatures achieved with $^3$He
refrigerators and conservative estimates of radiation loading for the
SPEED camera.\cite{CottinghamFSB2003}  While all detectors described
here have transitions close to the target value and transition widths
of $\sim$1~mK, the bolometer-to-bath thermal conductance is
unexpectedly high (factor of three) for detectors with Ti-Al-Ti leads
and within expectation for those with Mo leads.

\begin{figure}[htbp]
\centerline{\includegraphics{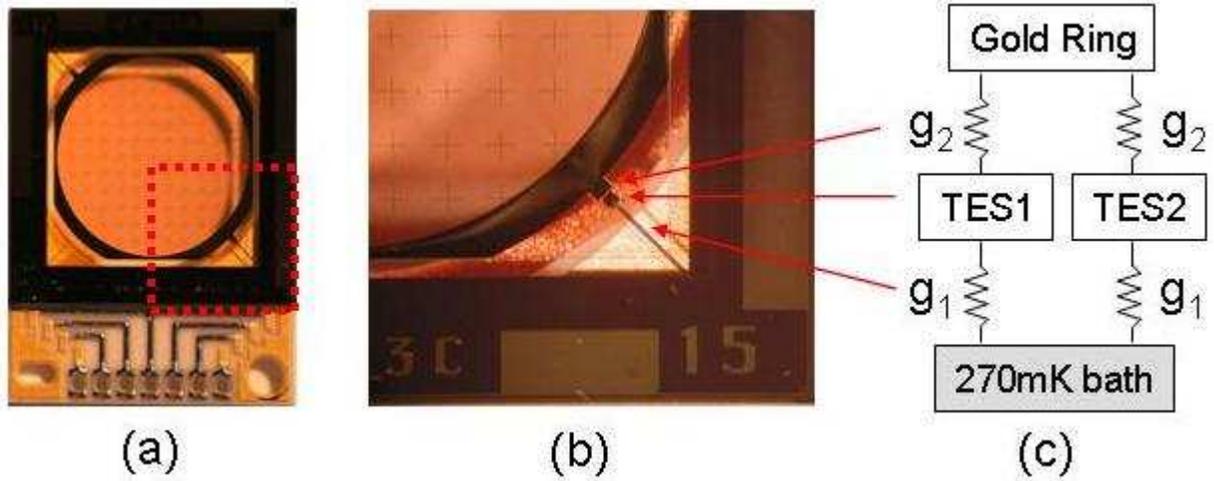}}
\caption{(a) Picture of complete FSB absorber frame.  (b) Close-up
showing TES and leads.  (c) Simple model describing the static thermal
behavior of detectors.  $g_1$ represents heat conductance from the TES
to the cold stage.  $g_2$ represents the gold-ring-to-TES thermal
conductance (see section \ref{g_model}).  The arrows indicate the
physical locations of elements in the schematic model and thermal
links between them.  $g_2$ corresponds to the small gap seen in (b)
between the TES and the ``stub'' extending off the gold
ring.}
\label{fsb_pics}
\end{figure}

\begin{figure}[htbp]
\centerline{\includegraphics{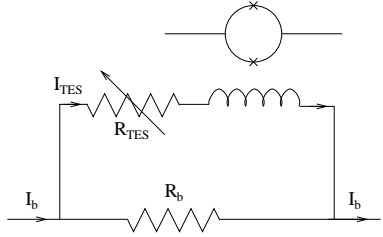}}
\caption{The TES is biased by running a current ($I_b$) through the
  shunt-TES combination.  The current flowing through the TES
  ($I_\mathrm{TES}$) is measured with a SQUID multiplexer.}
\label{TES_circuit}
\end{figure}

Due to the large absorbing area of FSBs and the poor thermal
conductance of the thin Si$_3$N$_4$ film, a 10~mm inner diameter,
150~nm thick gold ring is deposited around the absorbing area to
efficiently couple absorbing elements to the TES.  Thus, the absorbed
optical power is conducted through the gold ring to the two TESs and
then leaves the bolometer disk mainly through the TES leads.  This
simple model, adequate for describing the static behavior of
detectors, is illustrated in Fig.~\ref{fsb_pics}.c.  Despite its
inadequacies in describing dynamical behavior, we use this model as
most of the measurements presented here were conducted at very low
frequency.

The static thermal equation governing the biased TES is
\begin{eqnarray}
P + P_\mathrm{rad} = \int_{T_s}^{T} g_1(T') \mathrm{d}T',
\label{static_TES_eq}
\end{eqnarray}
where $P$ is the dissipated electrical power, $P_\mathrm{rad}$ is the
component of absorbed optical power ($Q$) reaching the biased TES, $T$
is the TES temperature, $T_s$ is the bath temperature, and $g_1(T)$ is
the thermal conductance between the TES and cold stage as indicated in
Fig.~\ref{fsb_pics}.c.  Because the TES resistance increases very
steeply with temperature, the negative feedback due to voltage biasing
keeps its temperature virtually constant within the transition,
especially in response to slow changes.  Because the right hand side of
Eq.~(\ref{static_TES_eq}) is virtually constant, a change in $P_\mathrm{rad}$
is essentially canceled by an opposite change in $P$, which we can
measure.  Assuming a constant temperature for the biased TES, the
measured power difference due to a change in absorbed optical power
$\Delta Q$ is
\begin{eqnarray}
\Delta P = - \Delta P_\mathrm{rad}  = - \frac{\Delta Q}{1 + \frac{g_1}{g_1 + g_2}}.
\label{delP}
\end{eqnarray}
Most of the absorbed power, therefore, flows through the biased TES if
$g_2 \gg g_1$ because its temperature is actively held constant.
Although Eq.~(\ref{delP}) is used in all our calculations, the
difference between $\Delta P_\mathrm{rad}$ and $\Delta Q$ is at most
20\% because $g_2$ is substantially larger than $g_1$ in the detectors
tested.


\section{Optical measurements}
\label{tests}

The absolute efficiencies of four FSB detectors from two fabrication
generations have been measured.  Each generation was tested separately
in stacks with a 271~GHz device followed by a 219~GHz device.  Each
detector stack was optically tested twice, once at The University of
Chicago and once at the University of Massachusetts Amherst.

\begin{figure}[htbp]
\centerline{\includegraphics{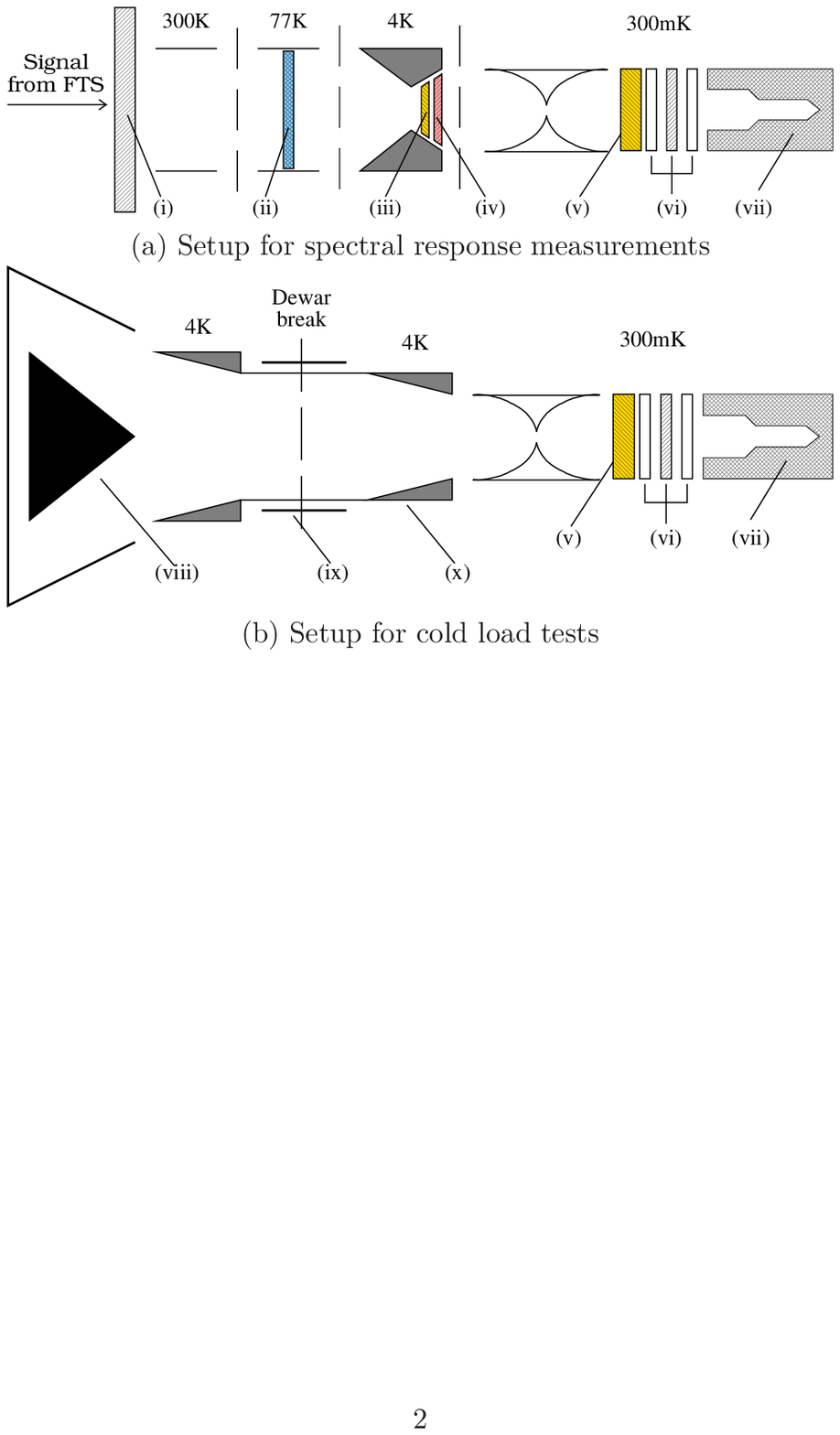}}
\caption{In both experimental arrangements light pipes are attached to
the several temperature stages of the cryostat and filters are
attached to the light pipes. (i) Polypropylene window ($\sim$0.25~mm)
at room temperature (ii) 1.3\% flat transmission Neutral Density
Filter (iii) 1.6~mm fluorogold (iv) 3.2~mm Teflon foam (Gore-tex) (v)
6.6~mm fluorogold (vi) 2 FSBs with a $\sim$6~mm gap (vii) Cavity made
of magnetically loaded epoxy to absorb transmitted radiation.  (viii)
Blackbody source capable of temperatures from 4-40~K housed in
separate cryostat with shared vacuum (ix) 77~K shielding around (x)
4.2~K light pipe.}
\label{fts_optics}
\end{figure}

Detectors are optically characterized in two steps: one step measures
the relative spectral response, and a second step measures the
absolute power absorbed due to a known source.  The absolute
absorption efficiency is the normalization factor multiplying the
relative spectral response that yields the measured absorbed power
from the known source.  The two optical setups are described below and
illustrated in Fig.~\ref{fts_optics}.

For both tests the optical setup at the detectors is the same.  The
detector absorber area is 10~mm in diameter.  Preceding the detectors
is a back-to-back Winston cone pair at 270~mK with a 10~mm diameter
and A$\Omega$ = 4.5~mm$^2$-sr.  This pair limits the radiation
incident upon the detector to an angle of 7.5$^\circ$ which, according
to simulations, does not significantly affect the detectors'
absorption spectra.  A $\sim$6~mm gap separates the two detectors.  To
prevent interference from reflected radiation a terminating absorber
with the shape shown in Fig.~\ref{fts_optics} is placed at the end of
the stack.  It is made from an absorptive epoxy similar to the coating
on the cold load described in Subsection 4.B.1.  Initial tests show
evidence of high frequency absorption by the detectors.  This is
expected \cite{kowitt96fsb} and proposed FSB based instruments have
provisions for optical low-pass filters ahead of the stack.  A 6.6~mm
piece of fluorogold\cite{duPont} was placed in the light path
following the Winston cone pair to absorb radiation above 750~GHz.
The system is cooled to sub-Kelvin temperatures with a closed cycle
$^3$He refrigerator.\cite{BhatiaChase00}

\subsection{Spectral response measurement}
\label{fts}

The spectral response measurements were conducted with a polarizing
Michelson Fourier Transform Spectrometer (FTS) originally built as a
prototype for the FIRAS experiment.\cite{fixsen94b}  A 1200~K
blackbody was used as the source of radiation.  In addition to
recording the spectrometer data, the blackbody was chopped at various
frequencies to establish the effective detector time constants.  While
the measured frequency responses are used to interpret spectrometer
data, all other measurements described here were conducted at
frequencies low enough to be unaffected by the thermal response time.
Fig.~\ref{fts_optics}.a shows the optical setup used for this set of
measurements.

The FSBs are designed for small radiative loads.  To reduce the
incident radiation to a level acceptable for the detectors, a neutral
density filter (NDF) with a flat transmission level of $\sim$1.3\%
over a broad frequency range is placed in front of the detectors.  The
incident radiation passes through a straight cone pair at 4.2~K that
acts as an A$\Omega$ restrictor and reduces the light pipe diameter to
10~mm.  To reduce the infrared radiative load on the cryostat, the
straight cone pair holds a thin piece of fluorogold and a thin piece
of Teflon foam\cite{Goretex}.  These low-pass
filters will marginally affect radiation at the two detector bands
primarily through the interference spectrum caused by internal and
external reflections at the filter surfaces (channel spectrum).


\subsection{Optical absorption measurement}
\label{eff}

\subsubsection{Setup}
\label{eff_setup}

To measure their response to changes in incident radiation, the
detectors were illuminated with a variable temperature black-body
source previously used for the same purpose in the TopHat
experiment.\cite{cordone04} The source consists of two concentric
cone-shaped cavities with internal walls shaped to enhance the number
of scatterings of incoming rays.  The internal walls are coated with a
1~mm thick layer of Eccosorb CR-114\cite{Eccosorb}, an absorber of
radiation within the band of interest.\cite{Hemmati1985} Ray trace
calculations using the source geometry and absorber properties show
that the source is black to within one part in $10^5$ over this
frequency range.  Therefore, we do not include an error due to its
non-unit emissivity.  The source is mounted on a separate cryostat
(cooled to 4.2~K) and connected to the detector cryostat with no
partitioning window and a common vacuum.  The light piping
configuration is illustrated in Fig.~\ref{fts_optics}.b.

The source temperature is controlled with heaters attached to its back
end.  The source is mechanically mounted on the 4.2~K plate by G-10
struts and thermally attached to it with two Pb straps.  The source
can be heated to 30~K without exhausting the liquid $^4$He bath too
quickly.  Source temperature modulations greater than 100~mK are
possible at frequencies below 0.2~Hz.  Two diode thermometers at
different ends of the source agree within their ratings and show no
phase difference during the modulations indicating that the source
is isothermal at these frequencies.  At these frequencies the
bolometer signal is also in phase with the temperatures (see
Fig.~\ref{ACtest}).

\subsubsection{Measurements}
\label{eff_measure}

\begin{figure}[htbp]
\centerline{\includegraphics{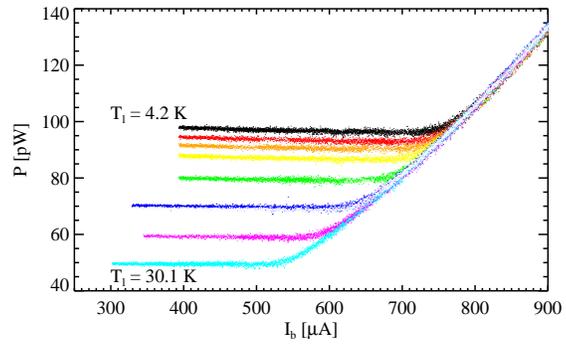}}
\caption{Electrical power dissipated in a 271~GHz detector at various source
temperatures from 4.2~K to 30.1~K.  The electrical power within the
transition (flat region) decreases with increasing source
temperature.}
\label{load_curves}
\end{figure}

\begin{figure}[htbp]
\centerline{\includegraphics{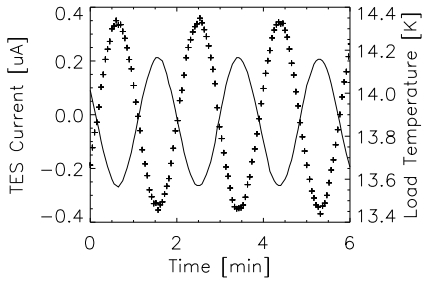}}
\caption{The solid curve is the measured source temperature modulation.
The plus signs are measured values of the TES current for a 271~GHz
detector.  At this slow modulation speed, the source temperature and
bolometer response are in phase (the TES current is AC coupled and
180$^\circ$ out of phase with bolometer loading).}
\label{ACtest}
\end{figure}

Changes in the power absorbed by each detector due to source temperature
variations are measured using two methods.  First, detector load curves
are made at several source temperatures by varying the bias current and
recording both the bias current and TES current (denoted $I_b$ and
$I_\mathrm{TES}$ respectively in Fig. \ref{TES_circuit}).  The bias current
is varied slowly so each data point represents an equilibrium
condition.  Fig.~\ref{load_curves} shows the electrical power
dissipated in the TES, computed from the measured currents, as a
function of bias current for several black-body temperatures.

The general shape of these curves can be understood through the static
thermal picture of Eq.~(\ref{static_TES_eq}) because the TES is in
equilibrium at every bias current.  The regions with flat electrical
power correspond to the TES being in its transition.  The power
changes very little because, with $P_\mathrm{rad}$ held constant (for each
curve), it simply reflects the small change in the r.h.s. of
Eq.~(\ref{static_TES_eq}) as the TES traverses the narrow transition.
The region where power increases quadratically is due to the TES
behaving as a pure resistor when normal ($P \simeq I_b^2 R_b^2 /
R_\mathrm{TES}$).  The superconducting region is not probed in these curves.

The flat regions assume different values for different source
temperatures because changes in $P$ must cancel changes in
$P_\mathrm{rad}$ at a given TES temperature.  We compute the changes
in absorbed radiation ($\Delta Q$) due to changes in source
temperature by measuring differences in electrical power ($\Delta P$)
and using Eq.~(\ref{delP}).  This method allows a measure of only
changes in optical power, not the absolute value of $Q$ at any
particular source temperature.

The electrical power ($P$) dissipated within the TES transition does
vary slightly due to the finite transition width.  Therefore we only
use data from a 5~m$\Omega$ span around a particular TES resistance
from each load curve.  If $R_\mathrm{TES}$ is uniquely determined by
$T$, selecting data in this way ensures that the r.h.s. of
Eq.~(\ref{static_TES_eq}) is constant over the data sets.  However,
$R_\mathrm{TES}$ has a small dependence on $I_\mathrm{TES}$.  Because
$R(T,I)$ is not well known for TESs, we do not correct for this effect
but estimate an error due to ignoring it by assuming that the chosen
resistance contour has the same shape as the critical current curve of
a superconductor on the $TI$ plane.

To solve for absorbed optical power differences $\Delta Q$ using
Eq.~(\ref{delP}), thermal conductances $g_1$ and $g_2$ must be known.
Based on the thermal conductivity of Si$_3$N$_4$ and the geometry,
$g_2$ is estimated at 21~nW/K with a large (50\%) uncertainty.  To
obtain $g_1$, the TES-to-stage thermal conductance was measured using
a separate set of load-curve data in which only the cold stage
temperature is varied between load curves.  The measured thermal
conductance, which includes both paths between TES~1 and the cold
stage in Fig.~\ref{fsb_pics}.c, is close to $2 g_1$ because $g_2$ is
substantially larger than $g_1$.  We find that the factor between
$\Delta P$ and $\Delta Q$ is only important for the detectors with
Ti-Al-Ti leads ($g_1$~$\simeq$~2.5~nW/K) and unimportant for the ones
with Mo leads ($g_1$~$\simeq$~0.7~nW/K).  Even for the former, $\Delta
P > 0.8 \Delta Q$.  Thus the large uncertainty in $g_2$ only
introduces a small error in the final results.

In the second method for measuring detector response to source
temperature changes, the source temperature was slowly modulated by
hundreds of mK with a constant bias applied to the TES.
Fig.~\ref{ACtest} shows both source temperature and TES current versus
time.  Eq.~(\ref{static_TES_eq}) when used to describe this equilibrium
situation yields as the detectors' static response
\begin{eqnarray}
\delta I = - \frac{\delta P_\mathrm{rad}}{I_b R_b} 
\frac{1}{
\frac{R_\mathrm{TES}-R_b}{R_\mathrm{TES}+R_b} + 
\frac{g_1(T)T}{P \alpha}\left(1 + \frac{\beta R_\mathrm{TES}}{R_\mathrm{TES}+R_b}\right)
}
\label{responsivity_eq}
\end{eqnarray}
where $\alpha = d \ln R_\mathrm{TES} / d \ln T $ and $\beta = d \ln
R_\mathrm{TES} / d \ln I$.

In the denominator of Eq.~(\ref{responsivity_eq}), the ratio of the
first term to the second is the loop gain of the electro-thermal
feedback.  As verified with load-curve data, this ratio is large for
the bias points used because each detector was biased within the lower
70\% of its transition.  Therefore, the second term in the denominator
is ignored when $\delta I$ is converted to $\delta P_\mathrm{rad}$ for use in
Eq.~(\ref{delP}) to evaluate $\delta Q$.  Typically, the fractional
error due to this simplification is $\sim$3\%.  Although the effects
of additional resistance in series with the TES are not included in
Eq.~(\ref{responsivity_eq}), we calculate an error due to the measured
sub-m$\Omega$ series resistance.


\section{Prototype optical characteristics}
\label{results}

\subsection{Spectral response}
\label{spect}

\begin{figure}[htbp]
\centerline{\includegraphics{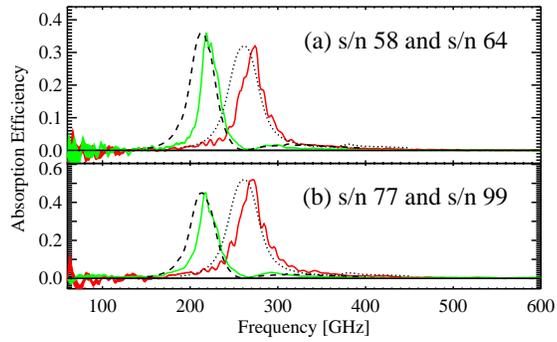}}
\caption{Solid lines show the spectral response of the two FSB stacks.
The difference between the two stacks is in the TES lead material
which should not affect optical characteristics.  In each stack, the
219~GHz detector is placed down stream of the 271~GHz detector in the
light path.  The line thickness of each curve represents the
uncertainty when it exceeds the constant line thickness used to
represent the data (mainly at low frequencies).  Overlaid with dashed
and dotted lines are the corresponding HFSS models scaled to have the
same peak absorption as the data.}
\label{fts_results}
\end{figure}

The shape of the radiation spectrum absorbed by a detector is measured
using the methods described in section~\ref{fts}.  To obtain the
relative spectral response of detectors we divide the measured profile
by a $\nu^2$ spectrum due to the $\sim$1200~K blackbody, cut off at
high-frequency by the fluorogold filters.  We ignore the channel
spectra of all filters when dividing out the incident spectrum due to
uncertainties in thickness and spacing of filters and because
unanticipated channel spectra, from gaps in light piping for example,
may be present in the incident radiation.  Fig.~\ref{fts_results}
shows the relative spectral response of detectors calculated in this
way.

The residual antisymmetric component of interferograms is used to
calculate the uncertainty in the absorption profiles and represented
in Fig.~\ref{fts_results} by the line thickness.  This uncertainty
accounts for the noise level at each frequency as well as systematics
due to the signal not being perfectly symmetrized.  The uncertainties
indicated in Fig.~\ref{fts_results} are small in the resonance regions
and only become significant at low frequencies due to $1/f$ noise.
Spectral shapes of Fig.~\ref{fts_results} are not completely accurate
representations of the intrinsic detector spectral response for two
reasons.  1) The channel spectra in the incident radiation have not
been divided out; the dominant ``ripple'' pattern seen on the curves
can be attributed to the thick (6.6~mm) fluorogold filter based on its
period.  However, the overall shapes of these curves may be affected
by longer period channel spectra from thinner filters.  But the
magnitude of such effects cannot be much greater than that of the
dominant channel spectrum from the thick fluorogold filter because the
relevant indices of refraction are no greater than that of fluorogold.
2) The notch in the incident spectrum due to the 271~GHz detector has
not been divided out of the 219~GHz detector's profile.  This effect
is clear in Fig.~\ref{fts_results} where a dip to near zero absorption
occurs in the 219 GHz detector at the resonance of the preceding
detector.

The small differences in shape between the two 271~GHz detectors are
on the order of the differences one would expect due to thin-filter
channel spectra dissimilarities between the two setups.  However, one
of the 219~GHz detectors has a $\sim$1\% (apparently uniform) level of
continuum absorption above the resonance, not seen in the other
similar detector.  We do not know the behavior of this continuum above
$\sim$800~GHz due to the low-pass filters in the setup.  We know that
it is not an analysis artifact because this continuum absorption shows
a dip at $\sim$560~GHz, expected due to water vapor absorption in the
spectrometer.  An inspection of detectors showed the
absorber-backshort separation was greater by about 15\% of the target
value in this device, possibly due to adhesive between the frames
peeling off, whereas the spacing was accurate to within 2\% in the
other FSBs.

Fig.~\ref{fts_results} also shows HFSS models of the two FSB channels.
The model curves have been scaled from their peak absorption near 80\%
to match the measured curves at the resonance peaks.  The measured
absorption profiles are slightly narrower than model predictions with Q
values ranging from 9 to 11, but are still well suited for use in
ground based instruments like the SPEED camera that utilize the mm-wave
transmission windows of the atmosphere.  The results described thus far
constrain only the relative optical absorption as a function of
frequency.  The next section explains how the absolute scale factor
for the curves of Fig.~\ref{fts_results} was established.

\subsection{Absolute absorption efficiency}
\label{abs_eff}

Under the assumption of a perfect black body, the measured response of
intervening filters, and the design A$\Omega$ of the Winston cone
pair, we calculate $B_I(\nu,T_l)$, the total amount of radiated power
from the source incident on the detector stack at a particular source
temperature $T_l$ as a function of frequency.  The amount of power
absorbed by the detector due to the source is then
\begin{eqnarray}
Q(\nu, T_l) = \epsilon \int_{0}^{\infty} f(\nu) B_I(\nu, T_l) d\nu.
\label{incident_calc}
\end{eqnarray}
Here $f(\nu)$ is the measured relative absorption profile of a
detector normalized so that its peak is at unity and $\epsilon$ is the
true fraction of incident power absorbed at the peak.

\begin{figure}[htbp]
\centerline{\includegraphics{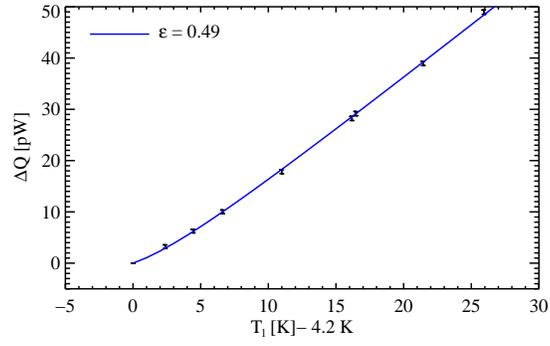}}
\caption{Estimate of absorbed optical power versus source temperature
  for a 271~GHz detector.  The curve through the data is the r.h.s. of
  Eq.~(\ref{incident_calc}) scaled to best fit the data.  The value of
  the best fit efficiency is also indicated.  The indicated error bars
  are statistical.  The error from the fit is combined with further
  systematics to get the final uncertainties of
  Table~\ref{results_table}.}
\label{dc_eff}
\end{figure}

Fig.~\ref{dc_eff} shows data from the load curve method described in
section \ref{eff_measure} along with a curve showing the
r.h.s. of Eq.~(\ref{incident_calc}) for the efficiency ($\epsilon$)
that best fits the data.  The data and calculated values are offset to
reflect only changes in power due to source temperature changes from
4.2~K.  Data from the source-temperature-modulation method are used
similarly to estimate efficiency.  Both methods yield good quality
fits to data within errors, as in Fig.~\ref{dc_eff}, indicating that
no major effects are being missed by this simple analysis.

The efficiencies for each detector obtained through both methods are
summarized in Table~\ref{results_table}.  While both methods were used
at both test facilities, the load curve (``DC tests'') results
presented here are obtained from one setup while the source modulation
(``AC tests'') results are taken from the other.  The general trend of
AC efficiencies being higher than DC efficiencies is due to a
difference between the two test facilities, not the difference in
method, because results of the two methods from each test setup are in
good agreement.  We show results from both setups to bring attention
to this trend which at the moment remains unexplained and is likely
due to differences in light piping and filter setups.

We also do not understand why the measured efficiencies ranging from
20\% to 60\% do not agree with the higher levels of $\sim$80\%
predicted by simulations.  We do not focus on this discrepancy because
we have not independently assessed the performance of models in
predicting absolute absorption efficiency.  On the other had, we did
expect the good agreement seen between models and measurement in terms
of spectral profiles, based on measured transmission spectra of single
FSB layers at room temperature.  The true resistivity of the gold on
the absorbing layer has not been measured.  A difference from the
ideal value (1.6~$\Omega$/square) by a factor of a few could also
explain the discrepancy.  We have also ignored waveguide-like effects
of the Winston horns treating them as pure A$\Omega$ limiters that
transmit 2.4 spatial modes at 219~GHz and and 3.7 modes at 271~GHz.
While the limiting approach angle to the FSBs of 7.5$^\circ$ causes an
insignificant deviation from the modeled absorption (according to
simulations), attenuation of spatial modes may result in losses as
high as 20\% given the 1.35~mm-diameter small orifice and the gradual
narrowing of the horn prior to it.  Because we ignore this effect,
formally, the absolute absorption efficiencies ($\epsilon$) quoted
here are for FSBs that couple to radiation through a 4.5~mm$^2$-sr
back-to-back Winston horn pair, a likely design for an instrument such
as the SPEED camera.

The error due to noise in the measurements (statistical error) is
small compared to estimates of systematic error.  Systematic errors
fall into two categories: errors in calculating the amount of
radiation power incident on detectors and errors in using detector
signals to estimate differences in absorbed power.  In calculating the
amount of incident optical power, uncertainties due to absolute and
differential thermometry errors of the source temperature are
included.  The transmission spectra of filters are necessary only in
the cases where different filters were used between the spectral
response and cold load tests.  Because there were such changes, we
measured the cold (4.2~K) transmission properties of the relevant
filters using a filter wheel setup to check the accuracy of their
transmission models.\cite{halpern86}  We also calculate an error due
to ignoring differences in channel spectra between the
spectral-response and cold-load measurements.  We attribute a further
10\% error in the incident optical power to non-idealities of the
light pipes and Winston cones.  While no account is made for changes
in optical power from sources other than the cold load during our
tests, we estimate the absorption due to such sources to be no more
$\sim$20~pW in most of our tests.

As for systematics in interpreting detector signals, the value and
standard deviation of a large sample of nominally 6~m$\Omega$ shunt
resistors ($R_b$) was established at 4.2~K and used in the analysis.
As described in section \ref{eff_measure}, we account for errors in
the relevant thermal conductances and for ignoring the current
dependence of TES resistance.  For the AC tests, we estimate an error
due to finite loop gain of the electro-thermal feedback.  We also
include errors due to possible misassignment of stray resistance in
series with the TES as on disk or off disk (i.e. whether the bolometer
is heated or not heated due to power dissipated through that
resistance).  A 1\% error in all gain factors in the cold and warm
electronics is also assumed.


\section{Conclusion}
\label{conclusion}

Tests of the prototype devices demonstrate that the FSB concept can be
realized with input from simulations to reliably set spectral
characteristics and that FSBs can be fabricated with reproducible
performance.  Because the detectors can be arranged in a stack with
high out-of-band transmission, individual channels have absolute
efficiencies comparable to the best seen in conventional bolometers
placed behind band defining filters.\cite{Turner2001,SCUBA1998} The
measured spectral profiles permit up to 8 simultaneous minimally
overlapping channels between 150-1500~GHz, a frequency range with much
potential for characterizing many interesting astronomical sources and
the backgrounds/foregrounds to those sources.  At present, there is
evidence that internal thermal couplings within detectors need
improvement.  More work is needed to understand the thermal and noise
properties of these detectors.  Another area of future research is the
use of FSBs with patterns of bars, not crosses, for polarization
sensitive measurements.

\section*{Acknowledgments}

This work is supported by NASA grant S20052896400000 and the Five
College Radio Astronomy Observatory under NSF grant AST-0228993. The
FSB devices were fabricated using the exceptional facilities of the
Detector Development Laboratory at NASA/GSFC.



\newpage

\begin{table}[h]
{\bf \caption{Physical parameters of FSBs.\label{FSBTable}}}\begin{center}
  \begin{tabular}{ccccc} \\ \hline
    $\nu_\mathrm{res}$ [GHz] & $l$ [$\mu$m] & $w$ [$\mu$m]& $g$ [$\mu$m] & s [$\mu$m]\\
	\hline
	219 & 572 & 14 & 1020 & 336 \\
	271 & 473 & 11.5 & 844 & 277 \\
	\hline 
  \end{tabular}
\end{center}
The columns, in order, are the center frequency ($\nu_\mathrm{res}$),
the length and width of a dipole element ($l$ and $w$), the
repetition scale of the cross pattern ($g$), and the spacing between
absorber and backshort ($s$).
\end{table}

\newpage

\begin{table}[h]
  {\bf \caption{The absolute absorption efficiency of the four tested
  FSBs.\label{results_table}}}\begin{center}
  \begin{tabular}{cccc} \\ \hline
    Detector & $\nu_c$ [GHz] & $\epsilon_\mathrm{DC}$ (\%) & $\epsilon_\mathrm{AC}$ (\%) \\ \hline
    s/n 58 & 219 & 28$\pm$6 & 44$\pm$4 \\
    s/n 64 & 271 & 24$\pm$5 & 39$\pm$4 \\
    s/n 77 & 219 & 40$\pm$5 & 50$\pm$5 \\
    s/n 99 & 271 & 49$\pm$7 & 55$\pm$5 \\ \hline
  \end{tabular}
\end{center}
The columns, in order, are the detector number, center frequency, the
absorption efficiency as measured with load-curve data, and absorption
efficiency as measured by modulating the source temperature.
\end{table}


\newpage

\section*{List of Figure Captions}

\noindent Fig. 1. Cartoon of a stack of two FSBs.  Of the radiation
entering from the left, the component in the band of FSB~1 (dotted
arrows) is absorbed and radiation outside that band passes through the
detector. Of the remaining light, radiation that is within the band of
FSB~2 (solid arrows) is absorbed and all other radiation (dashed
arrows) continues down the stack until absorbed by an additional FSB
or by a terminating absorber. On the right side is a cartoon
illustrating the combination of absorbing layer and reflecting
backshort.

\noindent Fig. 2. FSB surface geometry.

\noindent Fig. 3. (a) Picture of complete FSB absorber frame.  (b)
Close-up showing TES and leads.  (c) Simple model describing the
static thermal behavior of detectors.  $g_1$ represents heat
conductance from the TES to the cold stage.  $g_2$ represents the
gold-ring-to-TES thermal conductance (see section \ref{g_model}).  The
arrows indicate the physical locations of elements in the schematic
model and thermal links between them.  $g_2$ corresponds to the small
gap seen in (b) between the TES and the ``stub'' extending off the
gold ring.

\noindent Fig. 4. The TES is biased by running a current ($I_b$)
through the shunt-TES combination.  The current flowing through the
TES ($I_\mathrm{TES}$) is measured with a SQUID multiplexer.

\noindent Fig. 5. In both experimental arrangements light pipes are
attached to the several temperature stages of the cryostat and filters
are attached to the light pipes. (i) Polypropylene window
($\sim$0.25~mm) at room temperature (ii) 1.3\% flat transmission
Neutral Density Filter (iii) 1.6~mm fluorogold (iv) 3.2~mm Teflon foam
(Gore-tex) (v) 6.6~mm fluorogold (vi) 2 FSBs with a $\sim$6~mm gap
(vii) Cavity made of magnetically loaded epoxy to absorb transmitted
radiation.  (viii) Blackbody source capable of temperatures from
4-40~K housed in separate cryostat with shared vacuum (ix) 77~K
shielding around (x) 4.2~K light pipe.

\noindent Fig. 6. Electrical power dissipated in a 271~GHz detector at
various source temperatures from 4.2~K to 30.1~K.  The electrical
power within the transition (flat region) decreases with increasing
source temperature.

\noindent Fig. 7. The solid curve is the measured source temperature
modulation.  The plus signs are measured values of the TES current for
a 271~GHz detector.  At this slow modulation speed, the source
temperature and bolometer response are in phase (the TES current is AC
coupled and 180$^\circ$ out of phase with bolometer loading).

\noindent Fig. 8. Solid lines show the spectral response of the two
FSB stacks.  The difference between the two stacks is in the TES lead
material which should not affect optical characteristics.  In each
stack, the 219~GHz detector is placed down stream of the 271~GHz
detector in the light path.  The line thickness of each curve
represents the uncertainty when it exceeds the constant line thickness
used to represent the data (mainly at low frequencies).  Overlaid with
dashed and dotted lines are the corresponding HFSS models scaled to
have the same peak absorption as the data.

\noindent Fig. 9. Estimate of absorbed optical power versus source
temperature for a 271~GHz detector.  The curve through the data is the
r.h.s. of Eq.~(\ref{incident_calc}) scaled to best fit the data.  The
value of the best fit efficiency is also indicated.  The indicated
error bars are statistical.  The error from the fit is combined with
further systematics to get the final uncertainties of
Table~\ref{results_table}


\newpage

\newpage

\newpage

\newpage

\newpage

\newpage

\newpage

\newpage

\newpage

\end{document}